\documentclass[a4paper,11pt]{article}
\pdfoutput=1 

\usepackage{jcappub} 

\title{\boldmath Radial oscillations and stability of compact stars in $f(R, T) = R+ 2\beta T$ gravity}

\author[a]{Juan M. Z. Pretel, }
\author[a]{Sergio E. Jor\'as, }
\author[a,b]{Ribamar R. R. Reis}
\author[c,d]{and Jos\'e D. V. Arba\~nil}

\affiliation[a]{Instituto de F\'\i sica, Universidade Federal do Rio de Janeiro, \\ CEP 21941-972 Rio de Janeiro, RJ, Brazil}
\affiliation[b]{Observat\'orio do Valongo, Universidade Federal do Rio de Janeiro, \\ CEP 20080-090 Rio de Janeiro, RJ, Brazil}

\affiliation[c]{Departamento de Ciencias, Universidad Privada del Norte, \\ Avenida el Sol 461 San Juan de Lurigancho, 15434 Lima, Peru}
\affiliation[d]{Facultad de Ciencias F\'isicas, Universidad Nacional Mayor de San Marcos, \\ Avenida Venezuela s/n Cercado de Lima, 15081 Lima, Peru}

\emailAdd{juanzarate@if.ufrj.br}
\emailAdd{joras@if.ufrj.br}
\emailAdd{ribamar@if.ufrj.br}
\emailAdd{jose.arbanil@upn.pe}

\abstract{ We examine the static structure configurations and radial stability of compact stars within the context of $f(R, T)$ gravity, with $R$ and $T$ standing for the Ricci scalar and trace of the energy-momentum tensor, respectively. Considering the $f(R, T)=R+2\beta T$ functional form, with $\beta$ being a constant, we derive the corresponding hydrostatic equilibrium equation and the modified Chandrasekhar’s pulsation equation. The mass-radius relations and radial mode frequencies are obtained for some realistic equations of state. Our results show that the traditional stellar stability criteria, namely, the necessary condition $d M/d\rho_c >0$ and sufficient condition $\omega^2 >0$, still hold in this theory of gravity. }

\begin{document}
\maketitle
\flushbottom

\section{Introduction}\label{sec:level1}

Despite the great success of general relativity (GR) in predicting various phenomena that Newtonian gravitation fails to explain, nowadays, there are still unresolved issues. After the discovery of the accelerated expansion of the Universe, several extended theories of gravity have been developed; such as the so-called $f(R)$ theories \cite{SotiriouFaraoni, Felice}, $f(R,T)$ theories \cite{Harko2011}, among others that also include higher curvature corrections in the Einstein-Hilbert action. In the context of conventional GR, the well-known $\rm \Lambda$CDM model is based on cold dark matter and cosmological constant (which describes an unknown energy component). In principle, such a model gives a good agreement with the observational data, however, this model suffers from some problems that motivate the search for other possible theories of gravity. In fact, it is feasible to obtain accelerated expansion of the Universe in modified gravity without the need to introduce exotic forms of fluid like dark energy \cite{Nojiri2007}.  

Besides, standard GR is not renormalizable, so it cannot be quantized conventionally, as it is done in a quantum field theory. Indeed, the $S$ matrix of the theory of gravity developed by Einstein contains non-renormalizable ultraviolet divergences in four dimensions \cite{Goroff1986}. Nevertheless, it has been shown that higher-order actions can be renormalizable for appropriate coupling constants \cite{Stelle1977, Stelle1978}. Furthermore, these higher-derivative extensions are inevitable in the low-energy effective action of the string theory. Thus, considering higher-order curvature invariants to Einstein gravity can have important advantages.

In that regard, some important contributions have been made in the past. For a comprehensive review of modified theories of gravity and their cosmological consequences, see for instance Refs. \cite{Capozziello2011, Clifton2012, Nojiri2017} and references therein. On the other hand, at astrophysical scales, a broad overview about stellar structure models within the framework of extended theories of gravity formulated in both metric and metric-affine formalisms has recently been carried out in Ref. \cite{Olmo2020}.

In this paper, we will consider $f(R,T)$ gravity, which has been intensively studied in recent years. In fact, $f(R,T)$ gravity was phenomenologically introduced by  Refs. \cite{Dzhunushaliev2014, Yang2016, Liu2016} from the decomposition of the metric operator into classical and quantum parts. It is worth emphasizing that it is possible to construct models of $f(R,T)$ gravity consistent with the solar system experiments. In fact, the authors in Ref. \cite{Shabani2014} obtained the post-Newtonian gamma parameter for such theory, where it has been shown that this parameter depends on the cosmological matter too. Furthermore, on the galaxy scale, it has been shown that the mass corresponding to the interaction term (which appears in the modified field equations) leads to a flat rotation curve in the halo of galaxies \cite{Zaregonbadi2016}.

Over the last years, there has been a growing interest in constructing compact stars in the particular functional form we adopt in this paper, namely $f(R,T) = R+ 2\beta T$ gravity (proposed by Harko {\it et al.} \cite{Harko2011}) under the consideration of isotropic fluids \cite{Moraes2016, Das2016, Deb2018, Lobato2020} as well as with the inclusion of anisotropy \cite{Sharif2018, Deb2019, Maurya2019, MAURYA2020}. Nevertheless, the normal modes of the adiabatic radial vibrations in this theory have not yet been calculated in the literature. In that regard, the purpose of the present work is to investigate the stellar stability against radial pulsations in such theory of gravity for a sequence of equilibrium configurations with barotropic equations of state (EoS). To do so, we will follow a procedure analogous to that carried out by Chandrasekhar in GR \cite{Chandrasekhar}. Namely, we first consider stellar configurations by means of a spherically symmetric system composed of an isotropic perfect fluid, where the static background is described by the modified Tolman-Oppenheimer-Volkoff (TOV) equations. Our second step is to perturb all equations to first order and obtain the linearized equations for the radial oscillations. Finally, to verify if a given configuration is stable or unstable, we proceed to calculate the frequencies of its normal vibration modes. 

One of the advantages of using a function $f(R,T)$ linear in $R$, is that there is no extra degree of freedom as in non-linear theories where the Ricci scalar is a dynamical field and it does not vanish outside of a compact star. For instance, in the $f(R,T)= R+ 2\beta T$ model we have $R= 0$ outside the star and therefore it is possible to use the Schwarzschild solution \cite{Moraes2016, Das2016, Deb2018, Lobato2020}. Nonetheless, in $R$-squared gravity the Ricci scalar is obtained from a differential equation which arises from the trace of the field equations. In such theory the structure of compact stars is usually studied through non-perturbative \cite{Yazadjiev2014, PhysRevD.89.064019, Astashenok2015, 1512.05711, AstashenokOdinDom, Folomeev2018, Fulvio2020, Astashenok2020, Pretel2020JCAP} and perturvative \cite{Cooney2010, Arapoglu2011, Orellana2013, AstashenokCapOdi1} methods. It is worth commenting that the study of quasinormal modes of compact stars in $R^2$ gravity has been carried out by taking advantage of its mathematical equivalence with scalar-tensor theories \cite{Staykov2015, PhysRevD.98.104047, PhysRevD.98.044032, BlazquezSalcedo2019}, as well as in other models \cite{Mendes2018, Doneva2020, Aguiar2020, BlazquezSalcedo2020}. Furthermore, it has been investigated the structure of slowly \cite{Staykov2014} and rapidly \cite{Yazadjiev2015} rotating compact stars in $R$-squared gravity by using a non-perturbative and self-consistent method.

This paper is organized as follows: In Sec. \ref{sec:level2} we briefly summarize $f(R,T)$ gravity and we present the corresponding relativistic equations within the framework of $f(R,T) = R+ 2\beta T$ model. Section \ref{sec:level3} describes the static background for compact stars through modified TOV equations. In Sec. \ref{sec:level4} we derive the first-order differential equations that govern the radial oscillations in such model. In Sec. \ref{sec:level5} we present three well-known EoSs used in the literature to describe matter at high densities and pressures. Sec. \ref{sec:level6} presents a discussion of the numerical results for the equilibrium configurations as well as an analysis about their radial stability. Finally, our conclusions are summarized in Sec. \ref{sec:level7}. It is worth mentioning that throughout this work we use physical units and the signature $(-,+,+,+)$ will be adopted.


\section{$f(R, T)$ gravity}\label{sec:level2} 

Here we briefly summarize the $f(R, T)$ theories of gravity proposed by Harko {\it et al.} \cite{Harko2011}, where the modified form of the Einstein-Hilbert action is an arbitrary function of $R$ and $T$, the Ricci scalar and the trace of the energy-momentum tensor $T_{\mu\nu}$, respectively. Thus, in the presence of matter described by an action $S_m$, the full action is given by \cite{Harko2011}
\begin{equation}\label{1}
S = \frac{1}{2\kappa}\int d^4x\sqrt{-g}f(R,T) + S_m(g_{\mu\nu}, \Phi_m) , 
\end{equation} 
where $\kappa \equiv 8\pi G/c^4$ and the matter action depends on the metric $g_{\mu\nu}$ and the matter fields $\Phi_m$. The field equations in the metric formalism are obtained by varying the action (\ref{1}) with respect to the metric tensor, which yields
\begin{align}\label{2}
   f_R(R,T)R_{\mu\nu} &- \dfrac{1}{2}g_{\mu\nu}f(R,T) + [g_{\mu\nu}\square - \nabla_\mu\nabla_\nu] f_R(R,T) \nonumber  \\
   &= \kappa T_{\mu\nu} -f_T(R,T)T_{\mu\nu} - f_T(R,T)\Theta_{\mu\nu} ,
\end{align}
where, as usual, we have denoted $f_R \equiv \partial f(R,T)/\partial R$, $f_T \equiv \partial f(R,T)/\partial T$, $\square \equiv \nabla_\mu\nabla^\mu$ is the d'Alembert operator with $\nabla_\mu$ representing the covariant derivative, and the tensor $\Theta_{\mu\nu}$ is defined as 
\begin{equation}\label{3}
\Theta_{\mu\nu} \equiv g^{\alpha\beta}\frac{\delta T_{\alpha\beta}}{\delta g^{\mu\nu}} = -2T_{\mu\nu} + g_{\mu\nu}\mathcal{L}_m - 2g^{\alpha\beta} \frac{\partial^2\mathcal{L}_m}{\partial g^{\mu\nu} \partial g^{\alpha\beta}} ,
\end{equation}
with $\mathcal{L}_m$ being the matter Lagrangian density. 

As in $f(R)$ gravity \cite{SotiriouFaraoni, Felice}, in $f(R,T)$ theories the Ricci scalar is a dynamical quantity which is described by an equation obtained by taking the trace of the field equations (\ref{2})
\begin{equation}\label{4}
3\square f_R + Rf_R - 2f = \kappa T - (T+\Theta)f_T ,
\end{equation}
and the four-divergence leads to \cite{Barrientos2014}
\begin{equation}\label{5}
\nabla^\mu T_{\mu\nu} = \frac{f_T}{\kappa - f_T}\bigg[ (T_{\mu\nu} + \Theta_{\mu\nu})\nabla^\mu \ln f_T + \nabla^\mu\Theta_{\mu\nu} - \frac{1}{2}g_{\mu\nu}\nabla^\mu T \bigg] .
\end{equation}

In the present work we will focus on a particular class of $f(R,T)$ modified gravity models proposed in Ref. \cite{Harko2011}, namely, the $f(R, T) = R+ 2\beta T$ model, where $\beta$ is the only parameter of the theory. Such a parameter has to be restricted by observations from astrophysical up to cosmological scales. In that regard, some constraints on $\beta$ are currently being studied. For instance, in Ref. \cite{Lobato2020} is indicated that the neutron star crust is responsible for the fact that $\vert\beta\vert$ needs to be so small. From the dark energy density parameter, it was obtained $\beta \gtrsim -1.9 \times 10^{-8}$ \cite{Bhattacharjee2020}. From some observational data of massive white dwarfs, $\beta > -3.0 \times 10^{-4}$ \cite{Carvalho2017}. In addition, Ref. \cite{PhysRevD.95.123536} shows that the background evolution constraints $-0.1< \beta <1.5$ (at least), and Ref. \cite{Rudra2020} constraints a slightly more general limit (i.e., $f(R,T) = \alpha R^n + \beta T$), for which the best fit is $\beta = 0.8367$. As we will see in Section \ref{sec:level3}, it is already known that $\beta<0$.

In addition, as in Refs. \cite{Moraes2016, Das2016, Deb2018, Lobato2020}, for the matter Lagrangian corresponding to the fluid distribution, we will consider $\mathcal{L}_m =p$. Nevertheless, we must point out that there are also other choices for $\mathcal{L}_m$ in the literature \cite{Sharif2018}. Consequently, $\Theta_{\mu\nu} = -2T_{\mu\nu} + pg_{\mu\nu}$, $\Theta = -2T + 4p$, and Eqs. (\ref{2}), (\ref{4}), and (\ref{5}) take the following form
\begin{subequations}
\begin{align}
G_{\mu\nu} &= \kappa T_{\mu\nu} + \beta Tg_{\mu\nu} + 2\beta(T_{\mu\nu} - pg_{\mu\nu}) ,  \label{11a}  \\
R &= -\kappa T + 2\beta(-3T+4p) ,  \label{11b}  \\
\nabla^\mu T_{\mu\nu} &= \frac{2\beta}{\kappa + 2\beta}\left[ \nabla^\mu(pg_{\mu\nu}) - \frac{1}{2}g_{\mu\nu}\nabla^\mu T \right] ,   \label{11c}
\end{align}
\end{subequations}
where $G_{\mu\nu}$ is the Einstein tensor, and we recover equations corresponding to GR when $\beta = 0$. We point out that for $f(R,T)$ linear in $R$, as in the current paper, there is no extra degree of freedom and Eq.~(\ref{11b}) is an algebraic one, i.e., it is not a differential equation for $R$. In fact, in $f(R)= R+ \alpha R^2$ gravity, the Ricci scalar is also a dynamical field and $R \neq 0$ in the outer region of a compact star --- see Refs. \cite{Yazadjiev2014, PhysRevD.89.064019, Astashenok2015, 1512.05711, AstashenokOdinDom, Folomeev2018, Fulvio2020, Astashenok2020, Pretel2020JCAP} for discussions on the matter. 

In order to describe an isolated spherically symmetric compact star, we consider the line element
\begin{equation}\label{6}
ds^2 = -e^{2\psi}(dx^0)^2 + e^{2\lambda}dr^2 + r^2(d\theta^2 + \sin^2\theta d\phi^2) ,
\end{equation}
where $x^\mu = (ct, r, \theta, \phi)$ are the Schwarzschild  coordinates. The metric functions $\psi$ and $\lambda$ depend on both coordinates $x^0$ and $r$. 

On the other hand, the adiabatic and isotropic fluid that makes up the compact star is described by the energy-momentum tensor of a perfect fluid, which, in this work, it is given by
\begin{equation}\label{7}
T_{\mu\nu} = ( \epsilon + p )u_\mu u_\nu + pg_{\mu\nu} ,
\end{equation}
where $\epsilon= c^2\rho$ is the energy density ($\rho$ the mass density), $p$ represents the pressure of the isotropic fluid, and $u^\mu$ stands for the four-velocity of an observer comoving with the fluid, given by
\begin{equation}\label{8}
u^\mu = \dfrac{dx^\mu}{d\tau} = u^0\left( 1, \dfrac{dx^i}{dx^0} \right) ,
\end{equation}
with $\tau$ being the proper time.

For the case under consideration, only $u^0$ and $u^1$ are different from zero\footnote{We are considering a spherically symmetric system with motions, if any, only in the radial directions.}, so
\begin{equation}\label{9}
    T_2^{\ 2} = T_3^{\ 3} = p ,
\end{equation}
and the covariant divergence of expression (\ref{7}) provides
\begin{equation}\label{10}
    \nabla_\nu T_1^{\ \nu} = \partial_0 T_1^{\ 0} + \partial_r T_1^{\ 1} + \left[ T_1^{\ 1} - T_0^{\ 0} \right]\psi' + T_1^{\ 0}\left( \dot{\psi} + \dot{\lambda} \right) + \dfrac{2}{r}\left[ T_1^{\ 1} - p \right] ,
\end{equation}
where the overdots and primes stand for partial differentiation with respect to $x^0$ and $r$, respectively.
For the line element (\ref{6}) and since $T = -\epsilon + 3p$, the non-zero components of the field equations (\ref{11a}) are given by 
\begin{subequations}
\begin{align}
& \dfrac{1}{r^2}\partial_r(re^{-2\lambda}) - \dfrac{1}{r^2} = \kappa T_0^{\ 0} + \beta(-\epsilon + p + 2T_0^{\ 0}) ,  \label{12a}   \\  
& e^{-2\lambda}\left( \dfrac{2}{r}\psi' + \dfrac{1}{r^2} \right) - \dfrac{1}{r^2} = \kappa T_1^{\ 1} + \beta(-\epsilon + p + 2T_1^{\ 1}) ,  \label{12b}  \\  
& e^{-2\lambda}\left[ \psi'' + \psi'^2 - \psi'\lambda' + \dfrac{1}{r}(\psi' - \lambda') \right] + e^{-2\psi}\left[ \dot{\lambda}\dot{\psi} - \ddot{\lambda} - \dot{\lambda}^2 \right] = \kappa p + \beta(-\epsilon+ 3p) ,  \label{12c}   \\
& \dfrac{2}{r}e^{-2\lambda}\dot{\lambda} = \kappa T_0^{\ 1} + 2\beta T_0^{\ 1} ,   \label{12d} 
\end{align}
\end{subequations}
and using Eq. (\ref{10}), the non-conservation of the energy-momentum tensor (\ref{11c}) becomes
\begin{equation}\label{13}
   \partial_0 T_1^{\ 0} + \partial_r T_1^{\ 1} + T_1^{\ 0}(\dot{\psi} + \dot{\lambda}) + \left[T_1^{\ 1} - T_0^{\ 0}\right]\psi' + \dfrac{2}{r}\left[T_1^{\ 1} - p\right] = \frac{\beta}{\kappa + 2\beta}(\epsilon' - p') .
\end{equation}
Note that the conservation of $T_{\mu\nu}$ would be recovered in the particular case (not studied here) $\epsilon=p+{\rm constant}$, even if $\beta\neq 0$.


\section{Background and modified TOV equations}\label{sec:level3} 

For a stellar configuration in state of hydrostatic equilibrium none of the quantities depends on $x^0$ so that $u^\mu = (u^0, 0, 0, 0)$ from Eq. (\ref{8}). In other words, the  spacetime described by the metric (\ref{6}) is static and which implies that
\begin{align}\label{14}
    T_0^{\ 0} &= -\epsilon_0 ,   &    T_1^{\ 1} &= T_2^{\ 2} = T_3^{\ 3} = p_0 ,
\end{align}
where the subscript zero refers to the quantities that describe the equilibrium state. Thus, the integration of Eq. (\ref{12a}) leads to
\begin{equation}\label{15}
    e^{-2\lambda_0} = 1 - \frac{2Gm}{c^2r} .
\end{equation}

The relation (\ref{15}) allows us to characterize the mass within a sphere of radius $r$, given by
\begin{equation}\label{16}
m(r) = \frac{4\pi}{c^2}\int_0^r \bar{r}^2\epsilon_0(\bar{r})d\bar{r} + \frac{\beta c^2}{2G}\int_0^r \bar{r}^2[3\epsilon_0(\bar{r}) - p_0(\bar{r})]d\bar{r} ,
\end{equation}
so that at the surface, where the pressure vanishes, $m(r_{\rm sur}) \equiv M$ is the total mass of the star.

Consequently, from Eqs. (\ref{12b}), (\ref{13}), (\ref{15}) and (\ref{16}), the relativistic structure of an isotropic star in the state of hydrostatic equilibrium within the context of $f(R,T) = R+ 2\beta T$ gravity is described by the following modified TOV equations
\begin{subequations}
\begin{align}
\frac{dm}{dr} &= \frac{4\pi}{c^2}r^2\epsilon + \frac{\beta c^2}{2G}r^2(3\epsilon - p) ,  \label{17a}   \\
\frac{dp}{dr} &= -\frac{G(\epsilon + p)}{c^2(1+a)}\left[ \frac{m}{r^2} + \frac{4\pi}{c^2}rp - \frac{\beta c^2}{2G}r(\epsilon - 3p) \right] \left[ 1 - \frac{2Gm}{c^2r} \right]^{-1} + \frac{a}{1+a}\frac{d\epsilon}{dr} ,   \label{17b}  \\
\frac{d\psi}{dr} &= -\frac{1+a}{\epsilon + p}\frac{dp}{dr} + \frac{a}{\epsilon + p}\frac{d\epsilon}{dr} ,  \label{17c}
\end{align}
\end{subequations}
where we have defined $a \equiv \beta/(\kappa+ 2\beta)$, and the subscripts zero have been removed to avoid cluttering since all the quantities that appear in these equations correspond to the equilibrium state. The above system of equations reduces to the conventional TOV equations used in GR \cite{Tolman1939, Oppenheimer1939} when $\beta =0$, as expected.

We remark that for a barotropic EoS, i.e., $p = p(\epsilon)$, Eq. (\ref{17b}) can be rewritten as 
\begin{align}
\frac{dp}{dr} =& -\frac{G}{c^2}(\epsilon + p)\left[  \frac{m}{r^2} + \frac{4\pi}{c^2}rp - \frac{\beta c^2}{2G}r(\epsilon - 3p) \right] \left[ 1 + a\left( 1 - \frac{d\epsilon}{dp} \right) \right]^{-1} \left[ 1 - \frac{2Gm}{c^2r} \right]^{-1} ,
\end{align}
so that the configurations in hydrostatic equilibrium are obtained only when the following condition is respected
\begin{equation}
      1 + a\left( 1 - \frac{d\epsilon}{dp} \right) > 0,
\end{equation}
or alternatively
\begin{equation}
    a\left(-\frac{dp}{d\epsilon} + 1\right) < \frac{dp}{d\epsilon}, 
\end{equation}
and by taking into account that $dp/d\epsilon$ goes to zero at the surface of a compact star, we have $a<0$. Thus, as in the case of white dwarfs in $f(R,T)= R+ 2\beta T$ gravity \cite{Carvalho2017}, we will only consider negative values for $\beta$ in the present work.

Outside the compact star $\epsilon= p =0$ and, according to Eq. (\ref{11b}), we have $R \equiv 0 \, \forall \,r$, exactly as in GR,  due to the linearity of $f(R,T)$ in $R$, as mentioned before. This means that exterior spacetime is still being described by the Schwarzschild vacuum exterior solution, so that the continuity of the metric at the stellar surface (where $r= r_{\rm sur}$) imposes a boundary condition for Eq. (\ref{17c}). Therefore, given an EoS in the form $p= p(\epsilon)$ and ensuring regularity of the geometry at the center of the star, we solve the system of differential equations (\ref{17a})-(\ref{17c}) by taking into account the following boundary conditions
\begin{align}\label{18}
m(0) &= 0 ,   &   \epsilon(0) &= \epsilon_c ,   &   \psi(r_{\rm sur}) = \frac{1}{2}\ln \left[ 1 - \frac{2GM}{c^2r_{\rm sur}} \right] .
\end{align}


\section{Radial oscillations and stability equations}\label{sec:level4}

In order to study the radial stability of compact stars within the context of $f(R,T) = R+2\beta T$ gravity, it is necessary to calculate the frequencies of its normal oscillation modes. In that regard, we consider small deviations from the hydrostatic equilibrium where a fluid element located at $r$ is displaced to the radial coordinate $r+\xi(x^0,r)$ in the perturbed configuration. In other words, the stellar system described by Eqs. (\ref{17a})-(\ref{17c}) is subjected to small radial perturbations, without losing spherical symmetry, such that $h(x^0, r) = h_0(r) + \delta h(x^0,r)$, where the quantity $h$ stands for any metric or fluid variable and $\delta h$ is the \textit{Eulerian perturbation}. It should be note that the relation between Eulerian and Lagrangian perturbations is given by
\begin{equation}\label{19}
\Delta h(x^0, r) \equiv h[x^0, r+\xi(x^0, r)] - h_0(r) \cong \delta h + \frac{dh_0}{dr}\xi , \ \ 
\end{equation}
where $\Delta h$ is the \textit{Lagrangian perturbation}, that is, the change measured by an observer who moves with the fluid. 

The perturbation will cause motions in the radial directions ($u^1\neq 0$) so that, in the perturbed state, we can define $v \equiv dr/dx^0 = \partial\xi/\partial x^0$. Since the four-velocity $u^\mu$ must satisfy the normalization condition ($u_\mu u^\mu =-1$), it takes the form $u^\mu = (e^{-\psi}, e^{-\psi_0}v, 0, 0)$ to first order. Then, the non-vanishing components of the energy-momentum tensor (\ref{7}) are 
\begin{align}\label{20}
T_0^{\ 1} &= -(\epsilon_0+p_0)v ,   &   T_1^{\ 0} &= (\epsilon_0 + p_0)e^{2(\lambda_0-\psi_0)}v ,  \nonumber   \\
T_0^{\ 0} &= -\epsilon ,   &   T_1^{\ 1} &= T_2^{\ 2} = T_3^{\ 3} = p .
\end{align}

\subsection{Perturbed field equations}\label{subsec:level4.1}

Following the same approach developed by Chandrasekhar in GR \cite{Chandrasekhar}, here we shall neglect all quantities of second order and higher orders in the radial motions and we only preserve the linear terms. Thus, after perturbing the field equations (\ref{12a}), (\ref{12b}) and (\ref{12d}) together with the non-conservative Eq. (\ref{13}), we have 
\begin{subequations}
\begin{equation}\label{21a}
    \delta\lambda = -\frac{r}{2}(\kappa + 2\beta)(\epsilon_0 + p_0)e^{2\lambda_0}\xi = -(\psi'_0 + \lambda'_0)\xi ,
\end{equation}
\vspace{-0.8cm}

\begin{equation}\label{21b}
    \frac{\partial}{\partial r}(\delta\psi) = \left[ \frac{\kappa+ 3\beta}{\kappa+ 2\beta}\frac{\delta p}{\epsilon_0+ p_0} - \frac{\beta}{\kappa + 2\beta}\frac{\delta\epsilon}{\epsilon_0+ p_0} - \left( 2\psi'_0 + \frac{1}{r} \right)\xi \right](\psi'_0 + \lambda'_0) ,
\end{equation}
\vspace{-0.8cm}

\begin{align}\label{21c}
\delta\epsilon &= \frac{\beta}{\kappa + 3\beta}\delta p - \left( \frac{\kappa+2\beta}{\kappa+3\beta} \right)\frac{1}{r^2}\frac{\partial}{\partial r}\left[ (\epsilon_0+p_0)r^2\xi \right]  \nonumber  \\
&=  \frac{\beta}{\kappa + 3\beta}\Delta p -\xi\epsilon'_0 - \left( \frac{\kappa+2\beta}{\kappa+3\beta} \right)(\epsilon_0 + p_0)\frac{e^{\psi_0}}{r^2} \frac{\partial}{\partial r}\left( r^2\xi e^{-\psi_0} \right)  , 
\end{align}
\vspace{-0.8cm}

\begin{equation}\label{21d}
(\epsilon_0+p_0)e^{2\lambda_0 - 2\psi_0}\frac{\partial v}{\partial x^0} + \frac{\partial}{\partial r}(\delta p) + (p_0+\epsilon_0)\frac{\partial}{\partial r}(\delta\psi) + (\delta p + \delta\epsilon)\psi'_0 = \frac{\beta}{\kappa+2\beta} \frac{\partial}{\partial r}(\delta\epsilon - \delta p) .
\end{equation}
\end{subequations}
Furthermore, let us suppose that all perturbations have a harmonic time dependence of the form $\xi(x^0, r)= \chi(r)e^{i\sigma x^0}$ and $\delta h(x^0, r)= \delta h(r)e^{i\sigma x^0}$, with $c\sigma \equiv \omega$ being the characteristic frequency to be determined. Accordingly, using Eq. (\ref{21b}), the expression (\ref{21d}) becomes
\begin{align}
&\sigma^2(\epsilon_0+ p_0)e^{2\lambda_0- 2\psi_0}\chi = \frac{\kappa+ 3\beta}{\kappa +2\beta}\frac{d(\delta p)}{dr} - \frac{\beta}{\kappa + 2\beta}\frac{d(\delta\epsilon)}{dr} + \frac{\delta p}{\kappa + 2\beta}\left[ (2\kappa +5\beta)\psi'_0 + (\kappa+ 3\beta)\lambda'_0 \right]  \nonumber   \\
& \hspace{2.8cm} + \frac{\delta\epsilon}{\kappa +2\beta}[ (\kappa+\beta)\psi'_0 - \beta\lambda'_0 ] - (\epsilon_0 + p_0)\left( 2\psi'_0 + \frac{1}{r} \right)(\psi'_0+ \lambda'_0)\chi  .  \label{22}
\end{align}

For a barotropic EoS the pressure is a function only of the energy density, this is, $p= p(\epsilon)$. Then, by taking into account Eq. (\ref{21c}), we obtain 
\begin{align}\label{23}
\delta p =& -\chi p'_0 - \gamma p_0 \frac{e^{\psi_0}}{r^2}\frac{d}{dr}\left( r^2\chi e^{-\psi_0} \right) + \frac{\beta}{\kappa + 3\beta}\frac{dp}{d\epsilon} \left[ \Delta p + (\epsilon_0 +p_0)\frac{e^{\psi_0}}{r^2}\frac{d}{dr}\left( r^2\chi e^{-\psi_0} \right) \right], 
\end{align}
or alternatively,
\begin{equation}\label{24}
\Delta p = -\frac{\gamma p_0}{\mathcal{J}} \frac{e^{\psi_0}}{r^2}\frac{d}{dr}\left( r^2\chi e^{-\psi_0} \right) ,
\end{equation}
where $\gamma = \left( 1+ \epsilon/p \right)dp/d\epsilon$ is the adiabatic index at constant entropy, and $\mathcal{J}$ has been defined as
\begin{equation}\label{25}
    \mathcal{J} \equiv \frac{\kappa+ 3\beta}{\kappa+ 2\beta} \left( 1- \frac{\beta}{\kappa+ 3\beta}\frac{dp}{d\epsilon} \right) .
\end{equation}

\subsection{Equations for radial oscillations}\label{subsec:level4.2}

Once we have the expressions for the amplitudes of perturbations, the goal is to find the differential equations that describe the radial oscillations. Since all terms are now the amplitudes of the perturbations and quantities of the static background, we can delete all reference to subscripts zero. Since the calculation is a bit tedious, below we will only summarize the main steps to get the oscillation equations. After substituting Eq. (\ref{21c}) into (\ref{22}), we have
\begin{align}
\sigma^2(\epsilon+ p)e^{2\lambda- 2\psi}\chi =& \ \frac{1+ 2a}{1+ a}\left[ (\delta p)' + (2\psi'+ \lambda')\delta p \right] + \frac{1}{1+ a}\left[ a\mathcal{Q}' + \left( a\lambda' - \frac{\kappa+ \beta}{\kappa+ 2\beta}\psi' \right)\mathcal{Q} \right]   \nonumber   \\
&- (\epsilon + p)\left( 2\psi' + \frac{1}{r} \right)(\psi' + \lambda')\chi ,  \label{26}
\end{align}
with $\mathcal{Q}$ being given by
\begin{align}\label{27}
\mathcal{Q} &\equiv  \frac{d}{dr}[(\epsilon + p)\chi] + \frac{2}{r}(\epsilon + p)\chi  = \chi\epsilon' + a\chi(\epsilon' - p') - \left( \frac{\epsilon+ p}{\gamma p} \right)\mathcal{J} \Delta p .  
\end{align}

By means of Eqs. (\ref{17c}), (\ref{23}) and the $\theta\theta$-component of the field equations (\ref{12c}), the expression (\ref{26}) becomes
\begin{align}
\mathcal{G}(\Delta p)' =&\ \chi\left\lbrace \frac{1+a}{1+2a}(\epsilon+ p)\left[\sigma^2e^{-2\psi}- (\kappa+ 3\beta)p+ \beta \epsilon \right]e^{2\lambda} -\frac{1+3a}{1+2a}\frac{4}{r}p' + \frac{1-a}{1+2a}(\epsilon+ p)\psi'^2   \right.  \nonumber  \\
 & \left. + \frac{a}{1+2a}\left[ \frac{4}{r}\epsilon' + a(\epsilon'- p')\left( \frac{8}{r}+ \psi'+ \lambda' \right) -p'(\psi'+\lambda') - (\epsilon+ p)\left( \lambda'+ \frac{2}{r} \right)\psi' \right]  \right\rbrace  \nonumber  \\
 & -\Delta p \left\lbrace 2\psi'+\lambda' -\frac{a}{1+2a}\left[(\lambda'+ 3\psi')\frac{\epsilon+ p}{\gamma p}\mathcal{J} + \frac{d}{dr}\left( \frac{\epsilon+ p}{\gamma p}\mathcal{J} \right)\right] \right\rbrace  \nonumber \\
 & + \frac{a}{1+2a}\left[ a(\epsilon'- p')- p' \right]\chi' ,  \label{28}
\end{align}
where $\mathcal{G} \equiv 1- \displaystyle\frac{a}{1+2a}\left( \frac{\epsilon+ p}{\gamma p}\mathcal{J} \right)$.

Finally, if we introduce a new variable $\zeta \equiv \chi/r$, from Eqs. (\ref{24}) and (\ref{28}) we obtain the first-order time-independent equations governing the radial oscillations in $f(R,T) = R+2\beta T$ gravity, namely
\begin{align}
\frac{d\zeta}{dr} =& -\frac{1}{r}\left( 3\zeta+ \frac{\mathcal{J}}{\gamma p}\Delta p \right) + \psi'\zeta ,  \label{29}  \\
\mathcal{G}\frac{d(\Delta p)}{dr} =&\ \zeta\left\lbrace \frac{1+a}{1+2a}(\epsilon+ p)\left[\frac{\omega^2}{c^2}e^{-2\psi}- (\kappa+ 3\beta)p+ \beta \epsilon \right]re^{2\lambda} + \frac{1-a}{1+2a}(\epsilon+ p)r\psi'^2   \right.  \nonumber  \\
&\left. -\frac{1+3a}{1+2a}4p' + \frac{ar}{1+2a}\left[ \frac{4}{r}\epsilon' + a(\epsilon'- p')\left( \frac{9}{r}+ \psi'+ \lambda' \right) -p'\left(\frac{1}{r} +\psi'+\lambda'\right)  \right. \right.  \nonumber  \\
&\left.\left. - (\epsilon+ p)\left( \lambda'+ \frac{2}{r} \right)\psi' \right]  \right\rbrace + \frac{ar}{1+2a}\left[ a(\epsilon'- p')- p' \right]\zeta'   \nonumber \\ 
& -\Delta p \left\lbrace 2\psi'+\lambda' -\frac{a}{1+2a}\left[(\lambda'+ 3\psi')\frac{\epsilon+ p}{\gamma p}\mathcal{J} + \frac{d}{dr}\left( \frac{\epsilon+ p}{\gamma p}\mathcal{J} \right)\right] \right\rbrace .  \label{30}
\end{align}

It is evident that when $\beta = 0$ (which implies $a=0$, $\mathcal{G}=1$ and $\mathcal{J}= 1$), the system of equations (\ref{29}) and (\ref{30}) reduces to the pure GR case \cite{Gondek, Flores2010, Pretel2020}. Just as in Einstein's gravity, Eq. (\ref{29}) has a trivial coordinate singularity at the center ($r = 0$). In order that $d\zeta/dr$ to be finite everywhere, it is required that as $r \rightarrow 0$ the coefficient of $1/r$ term must vanish, so 
\begin{equation}\label{31}
    \Delta p = -3\frac{\gamma p\zeta}{\mathcal{J}}  \qquad \ \  \text{as}  \qquad  \ \  r\rightarrow 0 .
\end{equation}
On the other hand, the surface of the star is determined by the condition $p(r= r_{\rm sur})=0$. This implies that the Lagrangian perturbation of the pressure at the surface is zero, namely,
\begin{equation}\label{32}
    \Delta p = 0  \qquad \ \  \text{as}  \qquad  \ \  r\rightarrow r_{\rm sur} .
\end{equation}


\section{Equations of state}\label{sec:level5} 

In order to calculate some physical characteristic of compact stars such as the radius, mass, and the fundamental mode eigenfrequency, it is required to choose the micro-physical relation between energy density and fluid pressure. In this work, we use both strange quark matter and two nucleonic matter EoSs.

To investigate strange quark stars, MIT bag model EoS is usually employed. It describes a self-gravitating fluid composed by up, down, and strange quarks. This EoS is given by the simple relation
\begin{equation}\label{QEoS}
p = b(\epsilon - 4B) .
\end{equation}
The constant $b$ depends on both the chosen mass of the strange quark $m_s$ and QCD coupling constant. It usually varies from $b = 1/3$ for $m_s = 0$, to $b = 0.28$ for $m_s = 250\ \text{MeV}$. The bag constant $B$ lies in the range $0.982 < B < 1.525$ in units of $B_0 = 60\ \text{MeV}/\text{fm}^3$ \cite{Paschalidis2017}. In the present work, we will consider the particular case $b =0.28$ and $B = 1$.

On the other hand, to study neutron stars, the simplest composition of matter in neutron star cores is the $npe\mu$ (neutron-proton-electron-muon) composition. In this regard, in order to describe neutron stars in $f(R,T) =R+ 2\beta T$ gravity, we will consider the APR EoS \cite{APR1998} (also known in the literature as APR4 EoS and which hosts a three-nucleon potential and Argonne 18 potential with UIX potential) and the SLy EoS \cite{DouchinHaensel2001} (which is based on the SLy effective nucleon-nucleon interaction).


\section{Numerical results and discussion}\label{sec:level6} 

The stellar structure equations and the radial pulsation equations with their corresponding boundary conditions are integrated from the center to the surface of the star. To investigate the static equilibrium configurations, the stellar equilibrium equations are solved by using the fourth order Runge-Kutta method for a given central energy density and $\beta$. Once obtaining the coefficients of the radial pulsation equations, the radial stability equations are solved through the shooting method, which will be explained in more detail in subsection \ref{subsec:level6.2}. We point out that, at the GR limit for the radial oscillation modes, our code reproduces the results obtained in the literature \cite{VathChanmugam, KokkotasRuoff}.

\subsection{Equilibrium configurations}\label{subsec:level6.1}

\begin{figure}[t]
\centering 
\includegraphics[width=0.49\textwidth,origin=c]{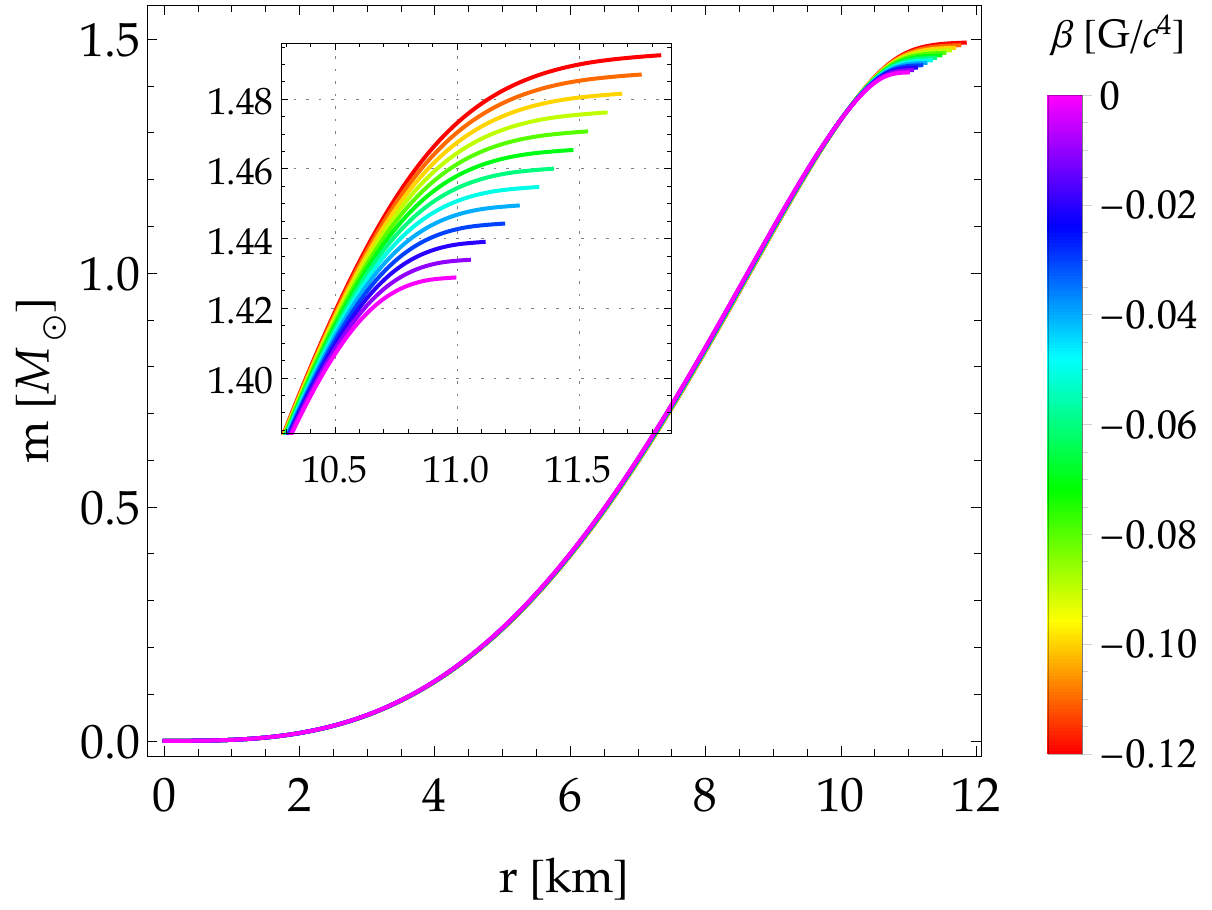} 
\includegraphics[width=0.49\textwidth,origin=c]{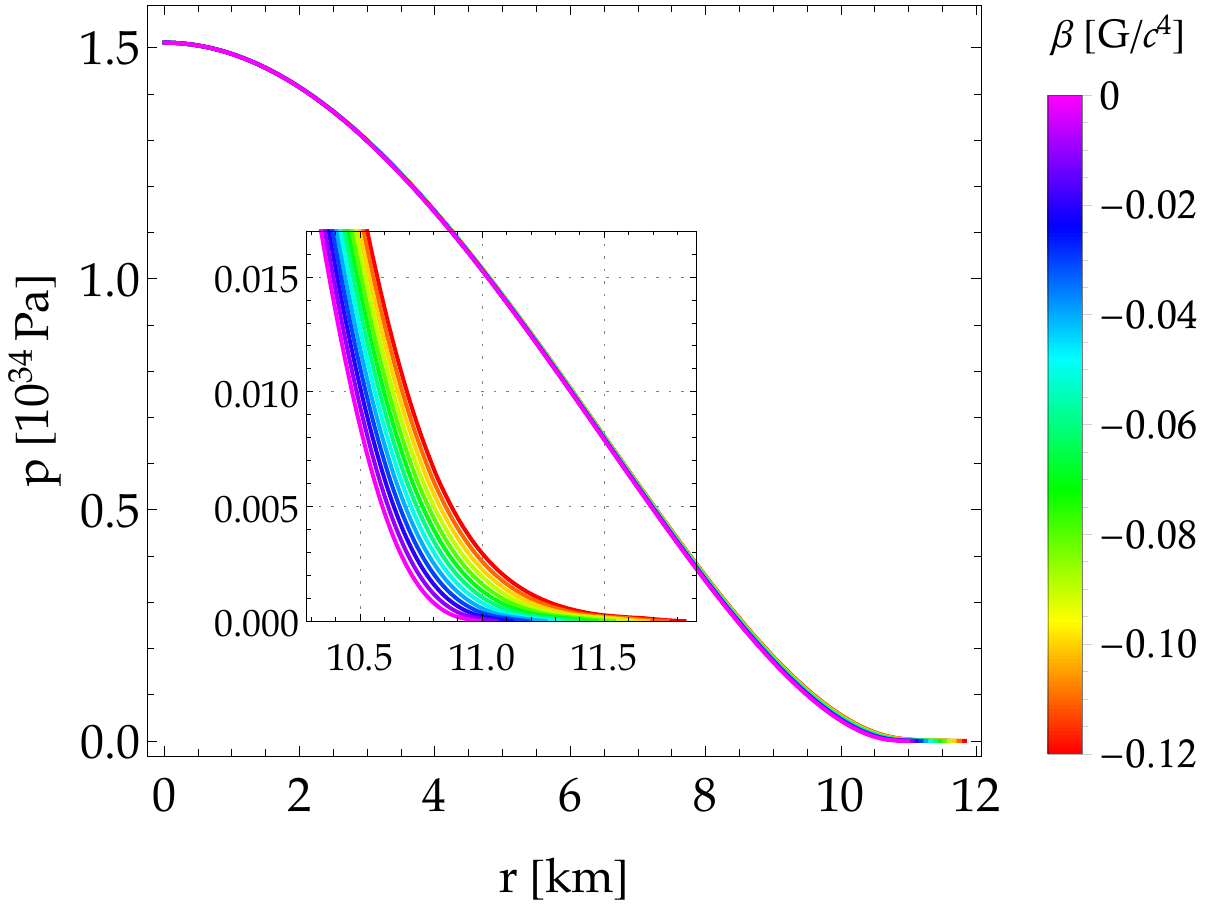}
\caption{\label{figure1} Radial behavior of the mass function (left panel) and pressure (right panel) for a central density $\rho_c = 1.0 \times 10^{18}\ \rm{kg}/\rm{m}^3$ with APR EoS and for different values of the parameter $\beta$. We can observe that the deviations from GR are only non negligible (but still far below any observational constraint) near the stellar surface. The more negative the value of $\beta$, the larger the radius of the neutron star. }
\end{figure}

Given an EoS and a specific value of the parameter $\beta$, the system of modified TOV equations (\ref{17a})-(\ref{17c}) with boundary conditions (\ref{18}) is numerically integrated from the stellar center to the surface $r= r_{\rm sur}$, which corresponds to a vanishing pressure. For instance, for a central density $\rho_c = 1.0 \times 10^{18}\ \rm{kg}/\rm{m}^3$ with APR EoS, Fig. \ref{figure1} illustrates the mass function and pressure as functions of the radial coordinate for different values of $\beta$. This indicates that the interior structure of a neutron star is modified by the parameter $\beta$ only near the surface and, as a consequence, the radius of the star increases as $\beta$ is more negative. 

\begin{figure}[t]
\centering 
\includegraphics[width=0.49\textwidth,origin=c]{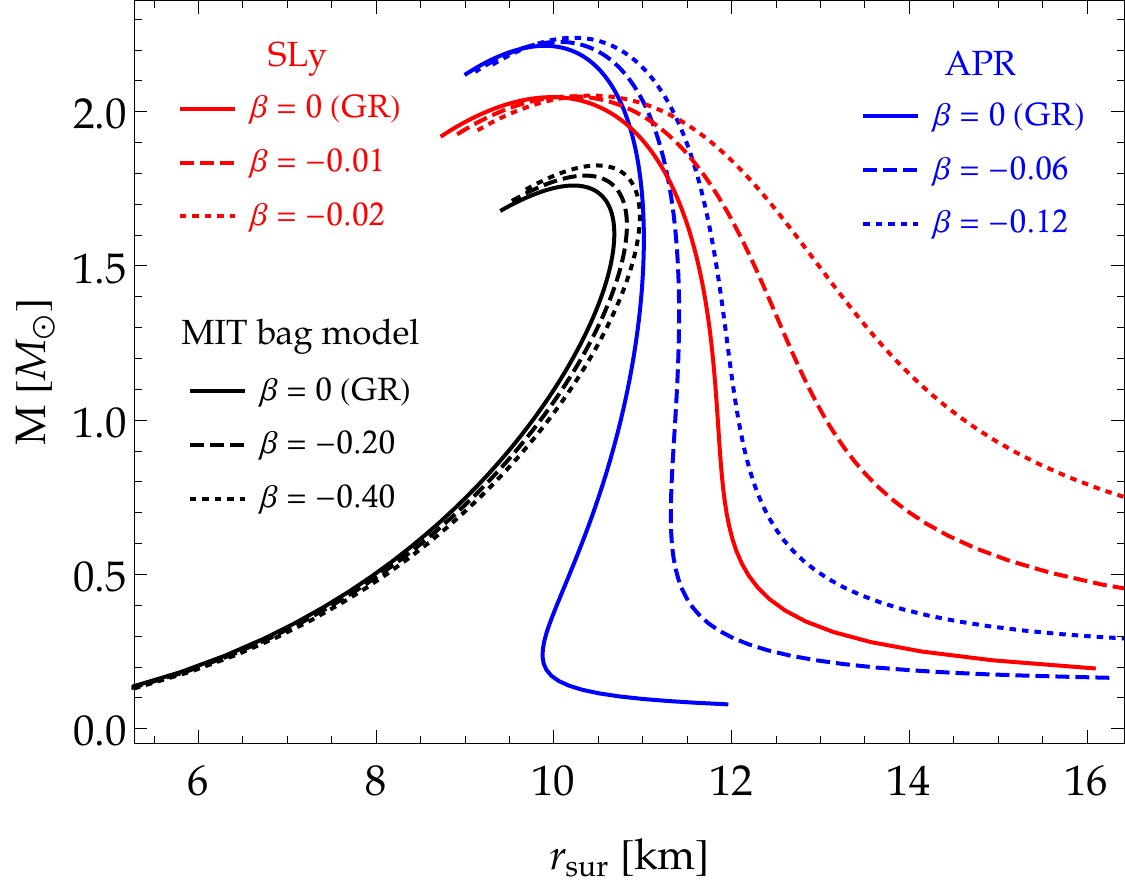} 
\includegraphics[width=0.49\textwidth,origin=c]{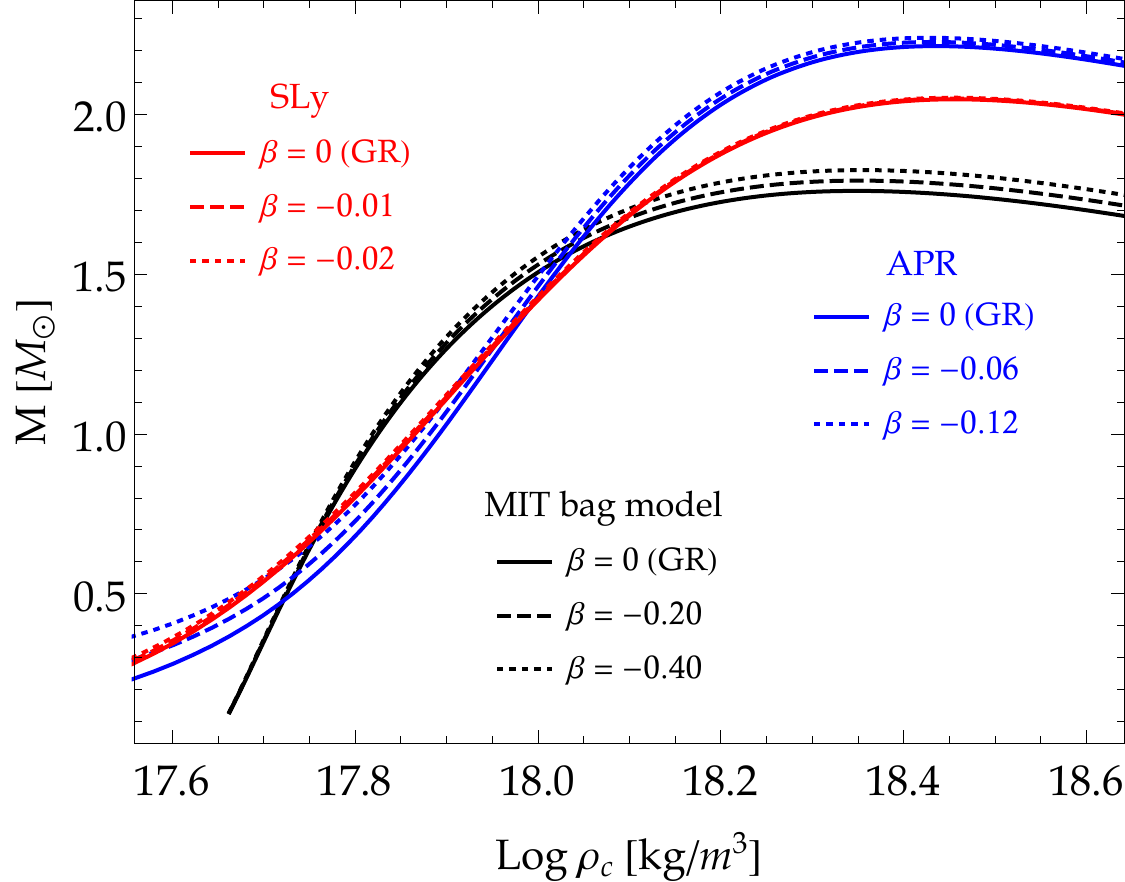}
\caption{\label{figure2} Mass-radius diagram (left panel) and mass-central density  relation (right panel) for compact stars in $f(R,T)= R+ 2\beta T$ gravity for different values of the parameter $\beta$ (given in $G/c^4$ units). The black, blue, and red lines correspond to the MIT bag model, APR, and SLy EoSs, respectively. In the case of neutron stars, it can be observed that the radius undergoes significant changes as we move away from GR for sufficiently low central densities. On the other hand, the radius suffers only subtle variations for quark stars. See also Fig.~\ref{figure3}. }
\end{figure}

\begin{figure}[t]
\centering 
\includegraphics[width=0.56\textwidth,origin=c]{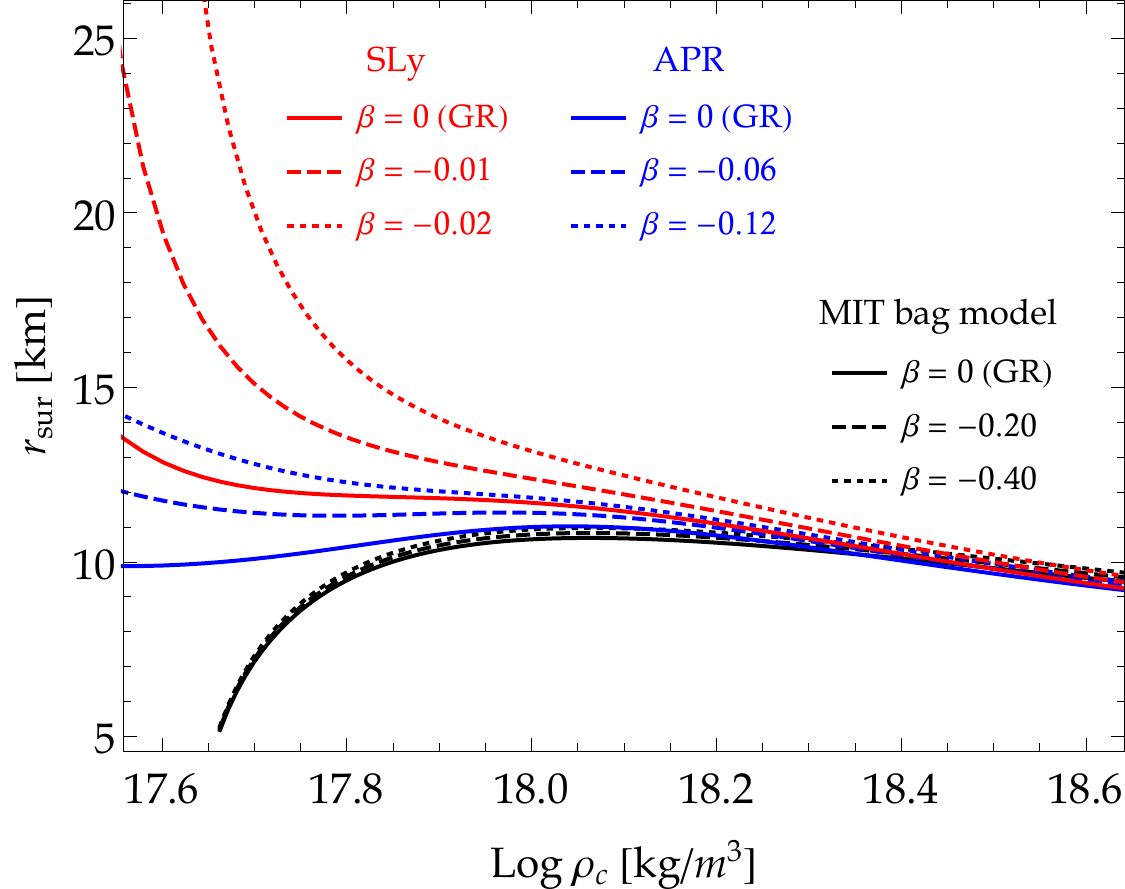} 
\caption{\label{figure3} Radius as a function of the central mass density for the equilibrium configurations presented in Fig. \ref{figure2}. For low central densities and SLy EoS, the radius diverges faster as $\beta$ is more negative. The same behavior happens for APR EoS, but for more negative values of $\beta$ (not shown). Meanwhile, for quark stars at low central densities the radius undergoes only slight modifications with respect to GR. }
\end{figure}

The total gravitational mass of a compact star is given by $M = m(r_{\rm sur})$. The mass-radius relations and mass-central density curves for quarks stars with EoS (\ref{QEoS}) and for neutron stars with APR and SLy EoSs in $f(R,T) = R+2\beta T$ gravity are presented in Fig. \ref{figure2}. The mass-radius diagram exhibits considerable deviations from GR for neutron stars only at sufficiently low central densities while in the higher central densities region (close to the maximum-mass point) the changes are smaller. 
Nevertheless, according to the right panel of Fig. \ref{figure2}, such theory of gravity does not significantly alter the masses (including their maximum values) of compact stars as functions of their central densities. On the other hand, a different behavior occurs for quark stars where the changes are irrelevant regardless of the value of $\beta$. This can be best observed on a plot of radius versus central density, as shown in Fig. \ref{figure3}, which shows a strong discrepancy between the models for low central densities. We conjecture that the MIT Bag model features almost no dependence on $\beta$ because, for this EoS, the ratio between the pressure gradient and the extra force $\propto \beta(\epsilon'- p')$ --- see Eq.~(\ref{13}) --- is constant through out the star (for any value of $\beta$ and for both high and low central densities). Meanwhile, for the other two EoSs, the extra force is increasingly larger than the pressure gradient as we approach the star surface --- and even more so for lower $\rho_c$, where the increase of the radius (compared to its GR value) is the largest.

The stellar configurations presented in Figs. \ref{figure2} and \ref{figure3} describe compact stars in hydrostatic equilibrium. Such equilibrium, however, can be either stable or unstable with respect to a small radial perturbation. In GR, it has been shown (see for instance Refs. \cite{Glendenning, Haensel2007}) that a turning point from stability to instability occurs when $dM/d\rho_c = 0$. This means that the stable branch in the sequence of stars is located before the critical density corresponding to the maximum mass. Thus, the stable stars on the $(\beta=0)$-curves in the right panel of Fig. \ref{figure2} are found in the region where  $dM/d\rho_c >0$. Due to its simplicity, this condition has been widely used in the literature even outside the scope of GR. Nevertheless, to the best of our knowledge, there is no formal proof of the equivalence between $dM/d\rho_c$ and stability in modified theories of gravity such as $f(R,T)$.
That is why in the next subsection we will analyze the stability from the calculation of the oscillation frequencies.

\begin{figure}[t]
\centering 
\includegraphics[width=0.50\textwidth,origin=c]{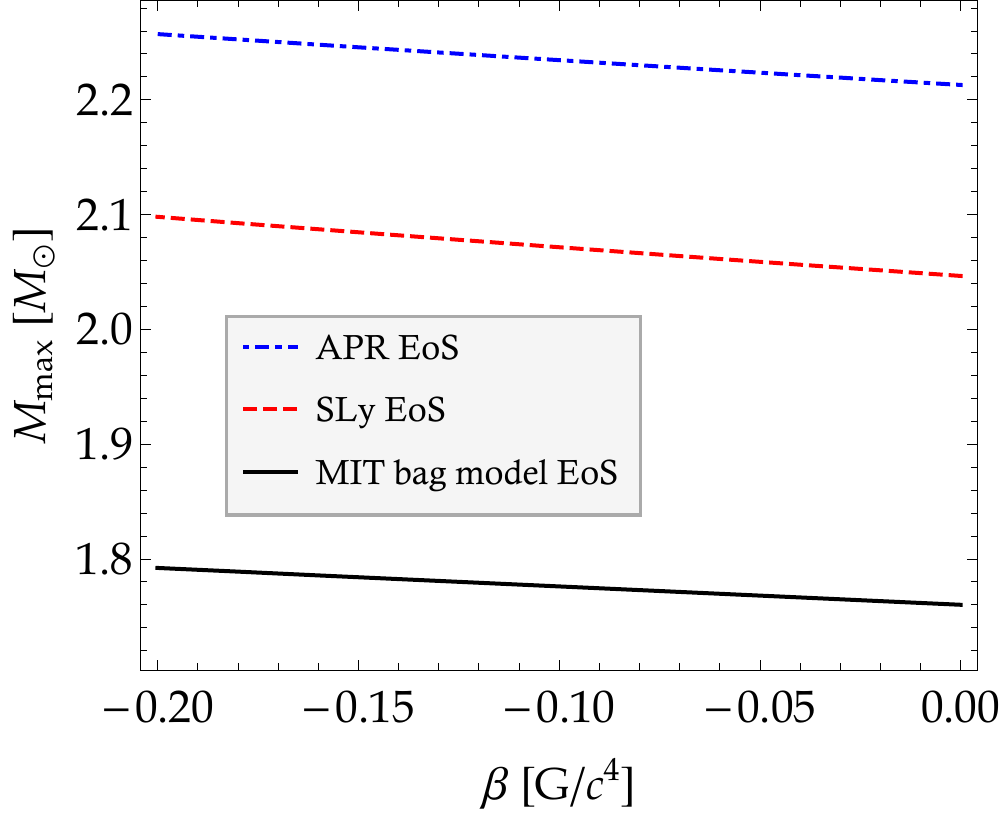} 
\caption{\label{figure4} Maximum mass as a function of the parameter $\beta$ for the three EoSs considered in Fig. \ref{figure2}. The same qualitative behavior occurs for each EoS and the maximum mass tends to its corresponding value in GR when $\beta \rightarrow 0$. }
\end{figure}

Before that, we briefly analyze how the maximum mass changes as we vary the parameter $\beta$ in the mass-radius relations. As illustrated in Fig. \ref{figure4}, the maximum mass has a linear behavior with beta regardless of the EoS, given by
\begin{equation}\label{MaxMass}
    M_{\rm max}= M_{\rm max, GR} - \mu\beta ,
\end{equation}
where $M_{\rm max, GR}$ is the maximum mass in GR, and $\mu$ assumes the following values in solar masses
\begin{equation}    
    \mu \approx \begin{cases}
        0.16 , \  \text{for MIT bag model EoS},  \\
        0.22 , \  \text{for APR EoS},  \\
        0.25 , \  \text{for SLy EoS}.  \\
    \end{cases}
\end{equation}
This is not a trivial result: For instance, for neutron stars in $f(R)= R+ \alpha R^2$ gravity, the maximum mass first decreases and after reaching a minimum it begins to increase monotonically with the parameter $\alpha$ \cite{Yazadjiev2014}.

\subsection{Radial pulsations and stability}\label{subsec:level6.2}

\begin{figure}[t]
\centering 
\includegraphics[width=0.49\textwidth,origin=c]{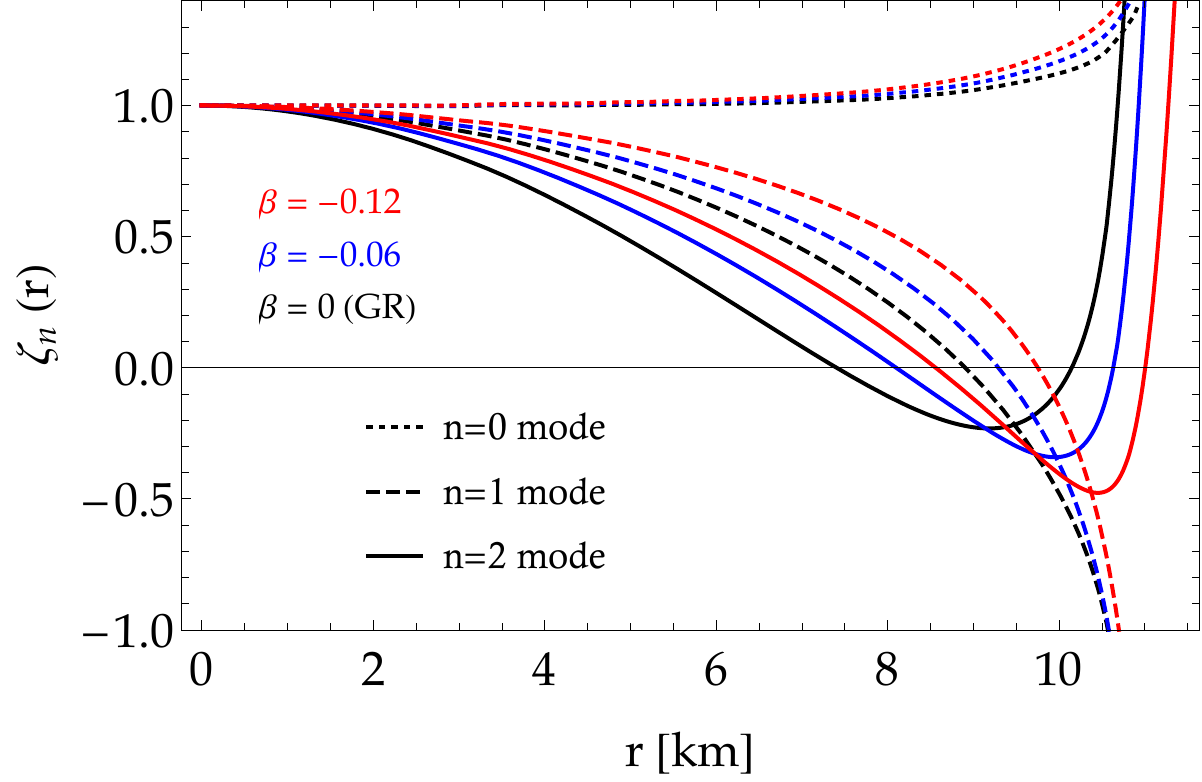} 
\includegraphics[width=0.49\textwidth,origin=c]{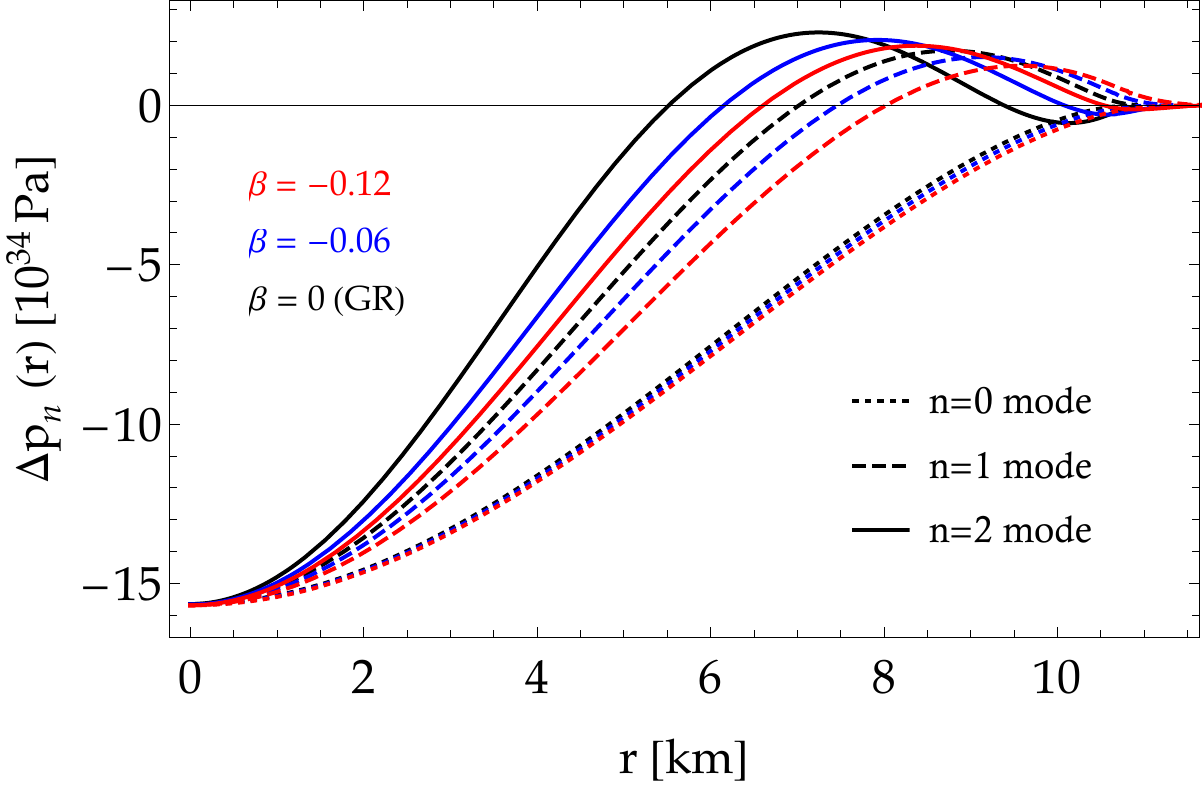}
\caption{\label{figure5} Numerical solution of the system of equations (\ref{29})-(\ref{30}) in the case of a neutron star with APR EoS and central density $\rho_{c} = 1.0 \times 10^{18}\ \text{kg}/\text{m}^3$ for three values of $\beta$. The different perturbations $\zeta_n (r)$ in the left panel and $\Delta p_n (r)$ in the right panel indicate the first three normal vibration modes as a function of the radial coordinate. The eigenfunctions $\zeta_n (r)$ have been normalized at the center of the star and the Lagrangian perturbations of the pressure $\Delta p_n (r)$ satisfy the boundary condition (\ref{32}) at the surface. }
\end{figure}

As in the pure GR case, once the quantities of the static background are computed through TOV equations, the numerical integration of first-order differential equations (\ref{29}) and (\ref{30}) is carried out using the shooting method, that is, we integrate the equations for a set of trial values $\omega^2$ satisfying the condition (\ref{31}). In addition, we consider that normalized eigenfunctions correspond to $\zeta(0) = 1$ at the origin, and we integrate to the surface. The appropriate frequencies of the radial pulsations correspond to the values for which the boundary condition (\ref{32}) is satisfied. In particular, for a central density $\rho_c = 1.0 \times 10^{18}\ \rm{kg}/\rm{m}^3$ with APR EoS and three different values of the parameter $\beta$, Fig. \ref{figure5} displays the perturbations $\zeta_n(r)$ and $\Delta p_n(r)$ for the first three eigenvalues $\omega_n^2$, where $n=0,1,2$ represents the number of nodes inside the star. In fact, the eigenvalue corresponding to $n=0$ is the fundamental mode, i.e., it has the lowest frequency and no nodes between the center and the surface of the star, whereas the first overtone $(n=1)$ has a node, the second overtone $(n=2)$ has two, and so on.

\begin{figure}[t]
\centering 
\includegraphics[width=0.49\textwidth,origin=c]{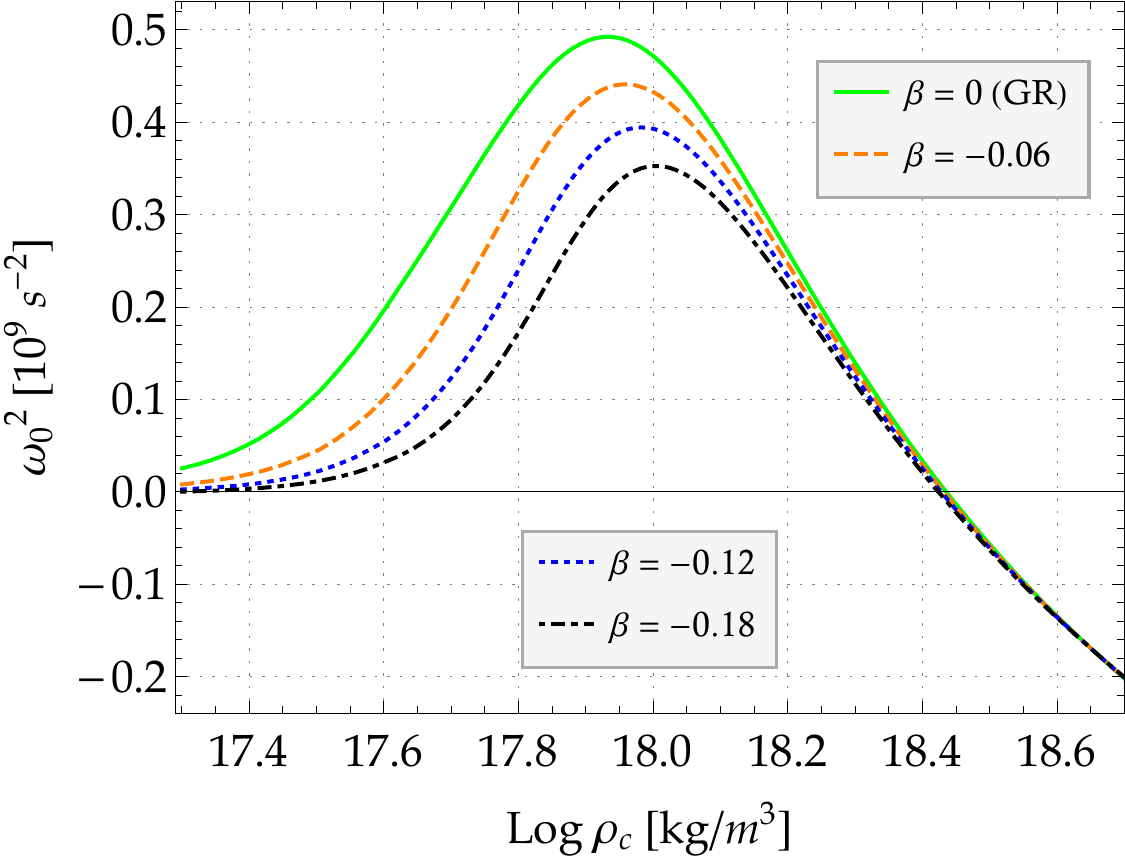} 
\includegraphics[width=0.49\textwidth,origin=c]{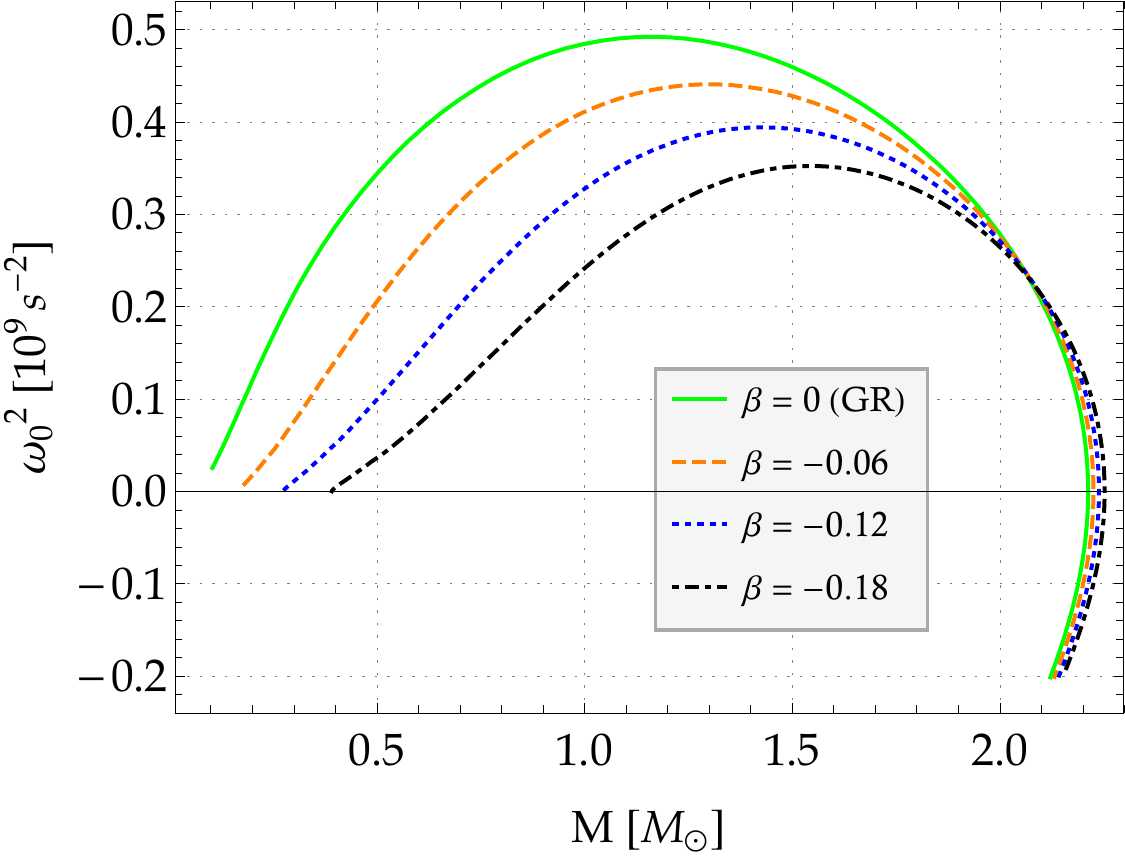}
\caption{\label{figure6} Left panel: Squared frequency of the fundamental oscillation mode versus central mass density predicted by the $f(R,T)= R+ 2\beta T$ model, where four values of the parameter $\beta$ are particularly considered for the APR EoS. Right panel: Squared frequency of the fundamental vibration mode as a function of mass. }
\end{figure}

Given an EoS and a value for the parameter $\beta$, we can now analyze the radial stability of the equilibrium configurations by calculating the oscillation frequencies. Figure \ref{figure6} shows the squared frequency of the fundamental oscillation mode against the central density (left panel) and total mass (right panel). According to the right plot presented in Fig. \ref{figure2} and the left plot in Fig. \ref{figure6}, the squared frequency of the fundamental mode passes through zero at the critical central density corresponding to the maximum-mass configuration. Therefore, regardless of the value of $\beta$, the maximum point on the curve $M(\rho_c)$ indicates the onset of instability for compact stars in $f(R,T)= R+ 2\beta T$ gravity. 

From Fig. \ref{figure6} we can also observe that the free parameter $\beta$ slightly modifies the radial stability of neutron stars. In other words, with the decrease of $\beta$, the onset of instability is indicated by a slightly smaller central density value. Furthermore, for low central densities the squared frequency of the fundamental mode goes asymptotically to zero as $\beta$ decreases.

\section{Conclusions}\label{sec:level7}

In this work we have studied the properties of compact stars in the $f(R,T)= R+ 2\beta T$ model, where the modified TOV equations have been integrated numerically to produce the mass-radius relations using some EoSs widely used in the literature. For neutron stars, the radii change significantly from GR, this is, for small masses the radius increases as $\beta$ is more negative whereas for large masses (near the maximum-mass point) the radius undergoes irrelevant changes. Meanwhile, for quark stars something different happens: for low densities both the mass and the radius do not change significantly, only at high densities the deviations are more pronounced with respect to Einstein's theory.

By perturbing the static background to first order in the metric and thermodynamic variables, we have derived for the first time the differential equations that describe the adiabatic radial oscillations in such a theory of gravity. We have analyzed the consequences of the extra term $2\beta T$ on radial pulsations, and our results show that the stellar stability criteria, traditionally used in GR, still hold in $f(R,T)= R+ 2\beta T$ gravity for quark stars and neutrons stars. This is, the necessary condition $dM/d\rho_c >0$ and the sufficient condition $\omega^2 >0$ are also fulfilled in $f(R,T)= R+ 2\beta T$ gravity.

The approach developed here for the radial perturbations can be applied to the stability analysis of compact stars with other equations of state for dense matter.

\acknowledgments

JMZP acknowledges Brazilian funding agency CAPES for PhD scholarship 331080/2019.


\end{document}